\documentclass[prd,aastex,amsmath,amssymb,twocolumn,floatfix,aps,showpacs]{revtex4-2}

\usepackage{amsmath}
\usepackage{amssymb}
\usepackage{textcomp}
\usepackage{xcolor}
\definecolor{myblue}{rgb}{0.0, 0.0, 0.6}
\usepackage{hyperref}
\hypersetup{
  colorlinks = true,
  citecolor  = myblue,
  linkcolor  = myblue,
  urlcolor   = myblue
}

\begin{document}

\title{
  Coulomb corrections for the non-flip and spin-flip electromagnetic $\boldsymbol{p}^\uparrow\!\boldsymbol{A}$ amplitudes
}

\author{A.~A.~Poblaguev}
%email{poblaguev@bnl.gov}
\affiliation{
 Brookhaven National Laboratory, Upton, New York 11973, USA
}

\begin{abstract}
  It is demonstrated that, within the eikonal approach, the Coulomb corrections to the elastic electromagnetic non-flip and spin-flip proton--nucleus amplitudes are identical when the two amplitudes share the same exponential form factors. This result allows Coulomb corrections to be computed numerically, and with high precision, for both electromagnetic and hadronic elastic $p^{\uparrow}A$ amplitudes in the massless-photon limit, including the effects of soft magnetic photon exchange. The method relies on analytical expressions and numerical integrations over a finite impact-parameter range with nonsingular integrands, providing a practical and systematically controlled framework for phenomenological applications.
\end{abstract}

\date{Fecruary 9, 2026}

\maketitle

In a recent publication \cite{Poblaguev:2025vtd}, it was proposed to derive the Coulomb correction to the non-flip elastic electromagnetic $p^\uparrow{A}$ amplitude $f_C(q^2)$ from the numerically calculated correction to the spin-flip amplitude $f_M(q^2)$. The latter can be evaluated with significantly less computational effort.

To validate this method, an analytical calculation in the leading order of the fine structure constant $\alpha$ \cite{Poblaguev:2021xkd}, as well as numerical calculations performed to all orders in $\alpha$ for $p^\uparrow{p}$ and $p^\uparrow\mathit{Au}$ scattering \cite{Poblaguev:2025vtd}, were considered. Although the disagreement between the calculated corrections to $f_C(q^2)$ and $f_M(q^2)$ was found to be small (see Figs.\,5 and 7 in Ref.\,\cite{Poblaguev:2025vtd}) within the momentum transfer range of interest, $0.001<q^2<0.02\,\text{GeV}^2$, the equality of the corrections was not demonstrated rigorously.
This equality is justified below.

Within the eikonal approach \cite{Franco:1973ei,*Cahn:1982nr,*Kopeliovich:2000ez}, the Coulomb-corrected amplitude $f_C^\gamma$ can be expressed in terms of its Born approximation,
\begin{equation}
  f_C(\boldsymbol{q}) = \frac{-2\alpha Z}{q^2}e^{-B_Cq^2/2},
\end{equation}
as
\begin{align}
  f_C^\gamma(\boldsymbol{q}) &= \frac{1}{2\pi}%
  \int{d^2\boldsymbol{b}\,e^{i\boldsymbol{q}\cdot\boldsymbol{b}}\,
    i\left[ 1 - e^{i\chi_C(\boldsymbol{b})} \right]
  }
  \nonumber \\
  &= \frac{-i}{2\pi}\int{d^2\boldsymbol{b}\,e^{i\boldsymbol{q}\cdot\boldsymbol{b}}\,
    \sum_{k=1}^\infty{\frac{\left(i\chi_C\right)^k}{k!} }
  }
  \label{eq:fCg} \\
  &= f_C(\boldsymbol{q})\times{\cal F}_C(q^2),
\end{align}
where
\begin{equation}
  \chi_C(\boldsymbol{b}) = \frac{1}{2\pi}%
  \int{d^2\boldsymbol{q}e^{-i\boldsymbol{q}\cdot\boldsymbol{b}} f_C(\boldsymbol{q}) }
  \label{eq:chiC}
\end{equation}
is the corresponding eikonal phase.

Notably, the inverse Fourier transformation of the eikonal term corresponding to the exchange of $m$ photons can be expressed as an integral involving $f_C(q^2)$,
\begin{align}
  F_C^{(m)}(\boldsymbol{q}) &=%
  \frac{1}{(2\pi)^2} \int{d^2\boldsymbol{b}e^{i\boldsymbol{q}\cdot\boldsymbol{b}}\,
    \chi_C^m(\boldsymbol{b}) }
  \nonumber \\
  &= \frac{1}{(2\pi)^2}\int{%
    d^2\boldsymbol{b}e^{i\boldsymbol{q}\cdot\boldsymbol{b}}\,%
    \int{
      \prod_{i=1}^m{\frac{d^2\boldsymbol{q}_i}{2\pi}%
            e^{-i\boldsymbol{q_i}\cdot\boldsymbol{b}}\,f_C(\boldsymbol{q}_i) }
      }
    }
  \nonumber \\
  &= \int{%
    \delta\left(\boldsymbol{q}-\sum_{i=0}^m{\boldsymbol{q}_i}\right)
    \prod_{i=1}^m{
      \frac{d^2\boldsymbol{q}_i}{2\pi}\,f_C(\boldsymbol{q}_i).
    }
  }.
  \label{eq:FCm}
\end{align}

For spin-flip electromagnetic elastic scattering, the Born approximation for the amplitude is given \cite{Vanzha:1972rps} by
\begin{equation}
  f_M(\boldsymbol{q})\bigg|_{B_M=B_C}% 
%  \frac{-\kappa_p\alpha Z}{2m_p}%
%  \frac{\kappa_p\,\boldsymbol{q}\cdot\boldsymbol{n}}{q^2}e^{-B_Mq^2/2}%
  = \frac{\kappa_p\,\boldsymbol{q}\cdot\boldsymbol{n}}{2m_p}%
  \times f_C(\boldsymbol{q}),
\end{equation}
where $\kappa_p=1.793$ is the anomalous magnetic moment of the proton, $m_p$ is the proton mass, $\boldsymbol{n}$ is a unit vector perpendicular to both the proton beam momentum and spin, and
it is assumed that the form factors of $f_M$ and $f_C$ are identical.
%it is assumed that form factors of $f_M$ and $f_C$ are the same.

To evaluate the Coulomb corrections to $f_M(q^2)$ corresponding to the term \eqref{eq:FCm}, one can replace, in all possible ways, only one non-flip amplitude $f_C(q_i^2)$ by the spin-flip amplitude $f_M(q_i^2)$,
\begin{align}
  F_C^{(m)}(\boldsymbol{q}) &\to%
  \int{\delta\left(\boldsymbol{q}-\sum_{i=0}^m{\boldsymbol{q}_i}\right) }\times
    \nonumber \\ &\qquad\qquad
     \sum_{j=1}^m{\left(
      \frac{\kappa_p\,\boldsymbol{q_j}\cdot\boldsymbol{n}}{2m_p}%
      \prod_{i}^m{\frac{d^2\boldsymbol{q}_i}{2\pi}\,f_C(\boldsymbol{q}_i) }
      \right) }
   \nonumber \\
   &= \frac{\kappa_p\,\boldsymbol{q}\cdot\boldsymbol{n}}{2m_p}%
   F_C^{(m)}(\boldsymbol{q}).
\end{align}
Applying this transformation to each term in Eq.\,\eqref{eq:fCg}, one arrives at
\begin{align}
  f_M^\gamma(\boldsymbol{q})\bigg|_{B_M=B_C} &=%
   \frac{\kappa_p\,\boldsymbol{q}\cdot\boldsymbol{n}}{2m_p}%
   f_C(\boldsymbol{q})\times{\cal F}_C(q^2)%
    \nonumber \\ &=
   f_M(\boldsymbol{q})\times{\cal F}_C(q^2)%
\end{align}
Since the Coulomb correction  ${\cal F}_M(q^2)$ to the spin-flip amplitude $f_M(\boldsymbol{q})$ can be expressed via known function $\widetilde{\cal F}(B_Cq^2/2,B_C/B_M)$ \cite{Poblaguev:2025vtd}, 
\begin{equation}
  {\cal F}_C(q^2) = {\cal F}_M(q^2)\bigg|_{B_M=B_C}  = \widetilde{\cal F}(B_Cq^2/2,1).
\end{equation}
The numerical calculation of the function $\widetilde{\cal F}(\widetilde{q}^2,\beta)$ is described in Appendix\,D of Ref.\,\cite{Poblaguev:2025vtd}.

Importantly, the adopted method provides a rigorous determination of the Coulomb correction to the non-flip electromagnetic amplitude without introducing a fictitious photon mass. Moreover, for a fixed value of the nuclear charge $Z$, the function $\widetilde{\cal F}(\tilde{q}^2,1)$ is a parameterless function of the dimensionless variable $\tilde{q}^2\!=\!B_Cq^2/2$. Therefore, $\widetilde{\cal F}(\tilde{q}^2,1)$ can be numerically pre-tabulated with the required precision, which can significantly simplify the analysis of experimental data, especially when $B_C$ is treated as a variable parameter.

Within this method, the following eikonal phase \cite{Poblaguev:2025vtd}
\begin{equation}
  \chi_C^\prime(b) = \alpha Z\left[\ln\left(\frac{b^2}{2B_C}\right) + E_1\left(\frac{b^2}{2B_C}\right)\right]
  \label{eq:chiC'}
\end{equation}
should be used for the evaluation of Coulomb corrections to other amplitudes (excluding ${\cal F}_C(q^2)$ and ${\cal F}_M(q^2)$). For example, for the hadronic non-flip amplitude,
\begin{align}
  f_N(q^2) &= (i+\rho)\frac{\sigma_\text{tot}}{4\pi}e^{-B_N q^2/2},
  \\
  \gamma_N(b) &= (i+\rho)\frac{\sigma_\text{tot}}{4\pi B_N}e^{-b^2/2B_N},
  \\
  f_N^\gamma(q^2) &= \int_0^\infty b\,db\,\gamma_N(b)\,e^{i\chi'_C(b)}J_0(bq),
  \label{eq:FgN}
\end{align}
where the logarithmic growth of $\chi_C^\prime(b)$ is inessential because the range of $\gamma_N(b)$ is Gaussianly confined by the value of the form-factor slope $B_N^{1/2}$.

The eikonal phase \eqref{eq:chiC'} also governs the modification of the Coulomb correction to $f_M(q^2)$ when $B_M\ne B_C$. From Eq.\,(D3) of Ref.\,\cite{Poblaguev:2025vtd}, one readily obtains
\begin{align}
  &{\cal F}_M(q^2) = {\cal F}_C(q^2)%
  \nonumber \\&\;\quad+%
  e^{-\tilde{q}^2}\tilde{q}\int{db%
    \left(e^{\frac{-b^2}{4}} - e^{\frac{-b^2}{4}\frac{B_C}{B_M}} \right)%
    e^{i\chi_C^\prime(b)} J_1\left(\tilde{q}b\right)}.
  \label{eq:FgMcorr}
\end{align}

Since the integrands in Eqs.\,\eqref{eq:FgN} and \eqref{eq:FgMcorr} are slowly varying functions of the impact parameter $b$ and are non-zero only within a limited range of $b$, the numerical integrations can be performed straightforwardly with sufficient accuracy.

In general, a form factor $F(q^2)$ can be represented as the sum of an exponential term and a non-exponential correction,
\begin{equation}
  F(q^2) = \left[e^{-Bq^2/2}\right]_\text{exp}%
  + \left[ F(q^2)-e^{-Bq^2/2} \right]_\text{non-exp},
  \label{eq:FF}
\end{equation}
where
\begin{equation}
  \vspace{+1pt}
  B = \frac{-2\,d\ln{F(q^2)}}{dq^2}\bigg|_{q^2=0}.
  \label{eq:B0}
\end{equation}
Because the non-exponential part of $F(q^2)$ is proportional to $q^4$, the associated eikonal phase $\delta\chi(b)$ can be calculated, even for $f_C(q^2)$, in the standard way \eqref{eq:chiC} via Fourier transformation. The region of $b$ where $\delta\chi(b)$ is non-zero is constrained similarly to $\gamma_N(b)$. Therefore, the Coulomb correction to the non-exponential part of the amplitude can be evaluated using $\chi_C^\prime(b)$, analogously to Eq.\,\eqref{eq:FgN}. Further details are given in Ref.\,\cite{Poblaguev:2025vtd}.

The method adopted in Ref.\,\cite{Poblaguev:2025vtd} also considers corrections due to magnetic photon exchange. Such corrections, also referred to as absorption corrections \cite{Kopeliovich:2023xtu}, may be significant in the case of polarized proton scattering off a large-$Z$ nucleus. However, multiple magnetic photon exchange was not included. Nevertheless, it was demonstrated that the corresponding corrections are small, and a method for evaluating them was provided.

To summarize, if multiple magnetic photon exchange is neglected and the form factors $F(q^2)$ (not necessarily exponential) of all Born amplitudes are known, then, regardless of the value of $Z$, the Coulomb corrections for elastic $p^{\uparrow}A$ scattering can be calculated numerically in a straightforward way with high precision. The achievable accuracy is limited primarily by computational precision and by the applicability of the eikonal approach.

\section*{Acknowledgments}
This work is authored by an employee of Brookhaven Science Associates, LLC under Contract No.\,DE-SC0012704 with the U.S. Department of Energy.

\bibliographystyle{apsrev4-2}
%\bibliography{_NumericalCoulomb_Comment.bib}

\end{document}